\documentclass{INTERSPEECH2023}

\interspeechcameraready 

\usepackage{stfloats}
\usepackage{subfigure}
\title{Zero-Shot Automatic Pronunciation Assessment}
\name{Hongfu Liu, Mingqian Shi, Ye Wang}
\address{
  School of Computing, National University of Singapore, Singapore}
\email{\{hongfu,m-shi,wangye\}@comp.nus.edu.sg}

\begin{document}

\maketitle
 
\begin{abstract}
% Automatic pronunciation assessment is a crucial technology for facilitating computer-assisted language learning. Prior methods have relied on either annotated speech-text data to train Automatic Speech Recognition (ASR) models or annotated speech-score data to train regression models. Although Self-Supervised Learning (SSL) pre-trained acoustic models have shown strong adaptation abilities on downstream speech tasks, their ability for zero-shot learning has not been fully explored. In this work, we propose a novel zero-shot automatic pronunciation assessment approach based on the SSL pre-trained acoustic model, HuBERT. Our method involves encoding the waveform speech input and transforming them into corrupted sequences via a masking module. We then employ the Transformer encoder of HuBERT and apply k-means clustering to obtain token sequences. Finally, a scoring module is formulated to evaluate pronunciation by measuring the number of wrongly recovered tokens. Notably, our proposed method is unsupervised and training-free. Experimental results on speechocean762 demonstrate that the proposed method achieves comparable performance compared to supervised baselines and outperforms non-regression baselines in terms of Pearson Correlation Coefficient (PCC). Additionally, we analyze how masking strategies affect the performance of the pronunciation assessment. 

Automatic Pronunciation Assessment (APA) is vital for computer-assisted language learning. Prior methods rely on annotated speech-text data to train Automatic Speech Recognition (ASR) models or speech-score data to train regression models. In this work, we propose a novel zero-shot APA method based on the pre-trained acoustic model, HuBERT. Our method involves encoding speech input and corrupting them via a masking module. We then employ the Transformer encoder and apply k-means clustering to obtain token sequences. Finally, a scoring module is designed to measure the number of wrongly recovered tokens. Experimental results on speechocean762 demonstrate that the proposed method achieves comparable performance to supervised regression baselines and outperforms non-regression baselines in terms of Pearson Correlation Coefficient (PCC). Additionally, we analyze how masking strategies affect the performance of APA.

% Automatic pronunciation assessment is an essential task in the application of computer-assisted language learning. Pr approaches either take Goodness of Pronunciation(GoP) as the measurement or train a regression model using supervised data. Recent self-supervised learning (SSL) models have been proven to be effective on downstream tasks in the speech community. In this work, we propose a novel zero-shot automatic pronunciation assessment approach based on SSL pre-trained model, HuBERT. This method first converts input speech frames into token sequences using k-means clustering. Then sequences, where parts of tokens are masked, would be fed into frozen HuBERT. Finally, a scoring module is designed to evaluate the gap between the output of HuBERT and the input sequence. Our proposed method is unsupervised and training-free. Experimental results on Speechocean762 demonstrate the effectiveness of our approach in terms of Pearson Correlation Coefficient(PCC).      

\end{abstract}

\section{Introduction}

Learning a second language (L2) is a common requirement in bilingual or multilingual communities. However, L2 learners often struggle with achieving good proficiency in pronunciation. Computer-assisted pronunciation training (CAPT) is a notable application that enables language learners to effectively learn the pronunciation of new languages~\cite{capt1,capt2}. CAPT provides feedback containing evaluation results, which can be automatically generated based on pronunciation, facilitating L2 learners in adjusting their pronunciation for improvement. Therefore, providing an overall assessment of pronunciation automatically is one of the primary objectives of CAPT.

Automatic pronunciation assessment has been extensively investigated over a prolonged period. Existing pronunciation assessment methods are implemented in the supervised setting. These approaches involve the usage of collected speech data with text annotations for training ASR models. Then the evaluation can be conducted based on the recognition results of ASR models. Goodness of Pronunciation (GoP) is one of the most commonly used metrics, aiming to provide phoneme-level scores for a given utterance. GoP requires calculating the log-posterior probability for each reference phoneme based on the contextual information~\cite{phonegop2000,phonegop,improvedgop}. On the other hand, there is an alternative research line that involves using speech data from non-native speakers with pronunciation scores annotated by domain experts to train regression models.  Various features of speech data have been explored in this line, one of which is the phone-level features of speech~\cite{gopfeat,speechocean762}. To enhance regression performance, ~\cite{deeptransfer} propose to use deep features transferred from the acoustic models of ASR. Using speech representations of pre-trained acoustic models such as wav2vec 2.0 or HuBERT also contributes to improving the regression performance by fine-tuning~\cite{gopwav2vec,sslapa}. Furthermore, multi-aspect pronunciation assessment at multiple granularities~\cite{trans-1,gopt} has been explored with multi-task supervised learning. However, there is a lack of unsupervised assessment approaches in the literature. All current pronunciation assessment methods require supervised signals to obtain the evaluation results.

% Existing pronunciation assessment approaches are conducted in the supervised setting. With collected speech data from non-native speakers and pronunciation scores annotated by domain experts, regression models can be trained to predict pronunciation scores. Common-used features of speech data include Goodness of Pronunciation (GoP), deep features from acoustic models or pre-trained models. To the best of our knowledge, there is a lack of unsupervised assessment approaches in the literature. Although it is feasible to avoid training regression models by taking GoP as a measurement of pronunciation, this approach still relies on a well-performed Automatic Speech Recognition (ASR) system. In addition, training an effective ASR system usually needs a large amount of labelled speech data. 

Resource-efficient methods have been widely investigated for the low-resource scenario in the speech community~\cite{lowasr,lowtts}. Nevertheless, it remains challenging to evaluate the quality of pronunciation using few or no data samples. Recent advances in Self-Supervised Learning (SSL) pre-trained language models (PLMs) have demonstrated strong few-shot and zero-shot learning abilities in the natural language processing community~\cite{fewlearner,zerolearner} due to the knowledge acquired during the pre-training stage. PLMs are capable of performing downstream tasks via appropriate prompting with limited or even no data samples. However, the zero-shot ability has not been fully explored for SSL pre-trained acoustic models. This is because they learn at the acoustic level and it is challenging to learn linguistic representations from raw audio~\cite{zeroben,zerocha}, making it difficult to adapt them to downstream tasks without fine-tuning. While fine-tuning SSL pre-trained acoustic models with supervised data has been shown to be effective in automatic pronunciation assessment~\cite{gopwav2vec,sslapa}, zero-shot pronunciation assessment has yet to be explored. Nevertheless, the acoustic-level knowledge acquired by SSL pre-trained acoustic models presents a viable option for zero-shot pronunciation assessment based on the unlabelled speech data observed during pre-training.

% Large-scale pre-trained models have shown the strong ability of few-shot and zero-shot learning in the natural language processing community \cite{fewlearner,zerolearner} due to the learned knowledge during the pretraining stage. With few or even no data samples, the pre-trained language model is able to perform downstream tasks via proper prompting. However, the zero-shot ability of SSL pre-trained acoustic models has not been well-exploited so far because they learn at the acoustic level with little semantic information. As such, it is difficult to adapt them to downstream tasks without fine-tuning them. While fine-tuning SSL pre-trained acoustic models using supervised data has shown to be effective on automatic pronunciation assessment, zero-shot pronunciation assessment has never been explored. The acoustic-level information learned by SSL pre-trained acoustic models provides an alternative to zero-shot pronunciation assessment based on the seen unlabelled speech data during pre-training.

In this work, we propose a zero-shot pronunciation assessment approach that requires no annotated speech data. This is achieved by leveraging the SSL pre-trained acoustic model, HuBERT~\cite{hubert}, for conducting the masked token prediction task. Our method involves encoding the waveform speech input into frame sequences and transforming them into corrupted sequences via a masking module. We then employ the Transformer Encoder of HuBERT and apply k-means clustering to obtain tokens of frame sequences and recovered tokens of corrupted sequences. Finally, a scoring module is designed to evaluate the pronunciation of a given speech by measuring the number of wrongly recovered tokens. Our proposed method is unsupervised and requires no fine-tuning. We conduct experiments on the speechocean762 dataset~\cite{speechocean762}. The experimental results demonstrate that the proposed method achieves comparable performance compared to supervised baselines and outperforms non-regression baselines in terms of the Pearson Correlation Coefficient.

\section{Method}

\begin{figure*}[htbp]
  \centering
  \includegraphics[scale=0.4]{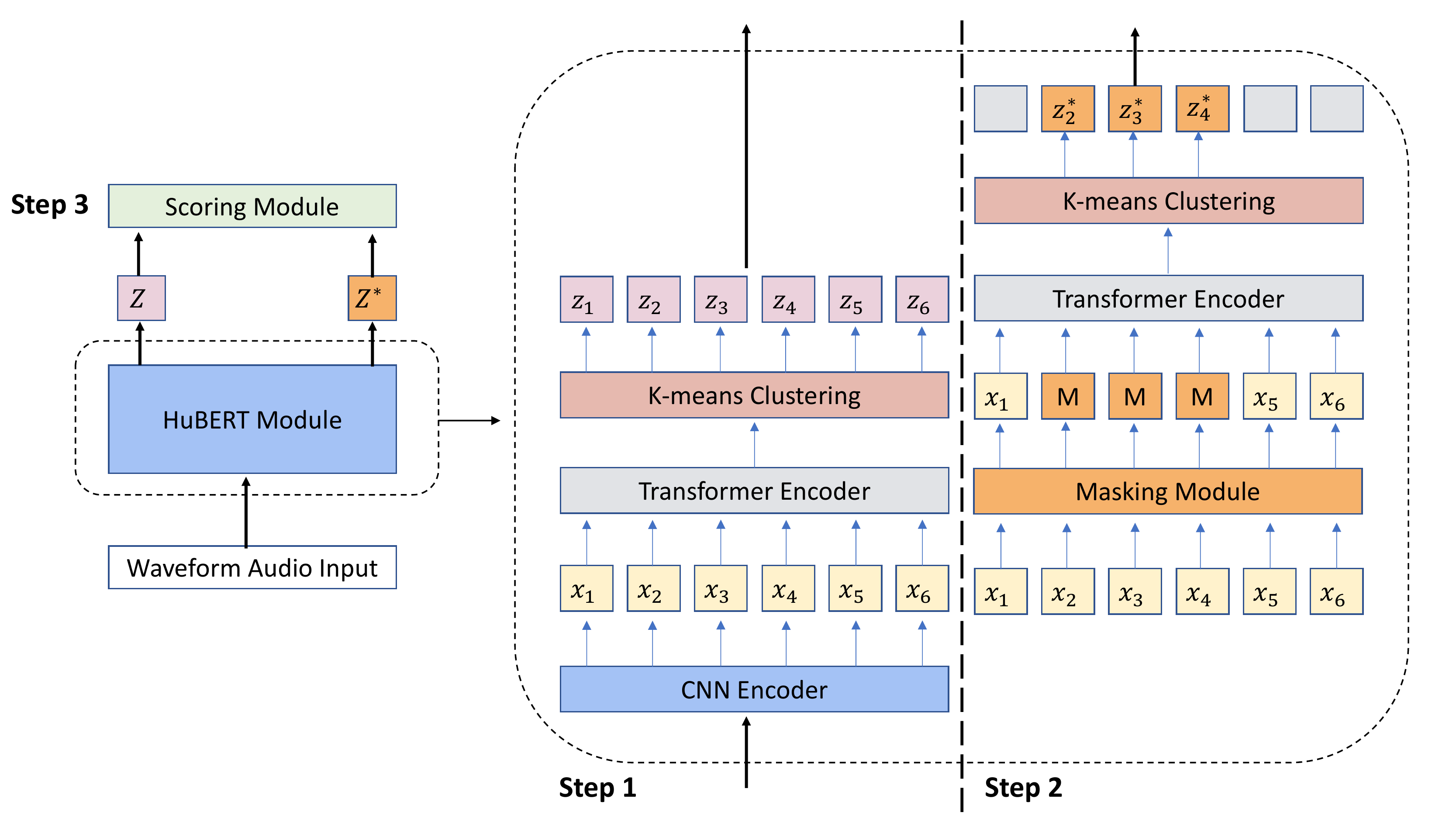}
  \caption{Overview of the zero-shot automatic pronunciation assessment.}
  \label{main_fig}
\end{figure*}

\subsection{Overview}
An overview of our proposed method is shown in Figure~\ref{main_fig}. We developed three main steps to achieve this assessment as shown in Figure~\ref{main_fig}. The first step is to input the waveform speech audio to the convolutional neural network (CNN) encoder to get a frame sequence. The Transformer encoder takes as input the frame sequence and the k-means clustering is utilized to obtain the token sequences. The second step is to apply a mask module on the frame sequence gained in Step 1 and input the masked sequence to the Transformer encoder followed by the k-means clustering to obtain the recovered tokens of masked spans. Finally, a scoring module is employed to measure the number of wrongly recovered tokens based on the outputs of Step 1 and Step 2. The intuition behind this is that for well-pronounced speech, recovered tokens would be similar to tokens of corresponding positions obtained from uncorrupted input. On the contrary, for mispronunciation speech, recovered tokens would differ from their counterparts a lot.

\subsection{HuBERT Module}
The HuBERT Module is adapted from the original HuBERT architecture~\cite{hubert}. This module consists of one CNN encoder, one Transformer encoder, one k-means clustering, and one masking module. Let $X = \{x_1, ..., x_T\}$ denote the output of the CNN encoder with $T$ frames. Then the Transformer encoder is employed to get latent representations of $X$ that are further utilized to obtain the token sequences $Z = \{z_1, ..., z_T\}$ through k-means clustering, where $z_t \in [C] $ is a $C$-class categorical variable and $t \in [T]$. $z_t$ is also known as the hidden acoustic unit.

\subsection{Masking Module}
To construct the masked token prediction task, we employ a masking strategy $r$ on $X$. If the set of indices to be masked is denoted by $M \subset [T]$ for a sequence $X$ with $T$ frames, then the corrupted version is denoted by $X^* = r(X, M)$, where $x_m$ is replaced by a mask embedding $x^*$ for $ m \in M $. Then we feed $X^*$ into the same Transformer encoder and use the same k-means clustering. As a consequence, the output token sequence of masked spans is denoted by $Z^* = \{z_m | m \in M\}$. 

Masking strategy is of great importance in the proposed method. Basically, we aim to mask mispronounced segments and expect the SSL pre-trained acoustic model can recover them with correctly pronounced tokens. However, whether the speech is mispronounced and where the mispronunciation occurs are unknown due to our unsupervised setting. To address this issue, we propose two strategies that are used to mask out the mispronunciation segments. 

\subsubsection{Random Masking}
Random masking is a direct approach that is based on the masking strategy employed in pre-training. However, a single instance of random masking may have a lower probability of covering the mispronunciation component. To address this concern, we propose to repeat random masking $k$ times for a given sequence $X$. Specifically, we randomly select $p\%$ of the frames in $X$ as starting indices, and subsequently mask spans of $l$ for each start index. These spans are mutually exclusive, with no overlap between them. By increasing the value of $k$, it is possible to ensure that each frame is masked at least once. 

\subsubsection{Regular Masking}
Regular masking is an alternative approach that masks frames in a rule-based way. This strategy involves segmenting the input into $k$ slices of equal length. We then proceed to mask one of those segments at a time and perform inference. The process is repeated until every segment has been masked at least once. The number $k$ of segmented slices determines the granularity of the segmentation. 
% Larger $k$ indicates fine-grained segmentation with shorter segments.

\subsection{Scoring Module}
In order to assess the quality of speech pronunciation, we introduce the scoring module, which measures the number of incorrectly recovered tokens based on $Z$ and $Z^*$. Specifically, the average Mis-Recovered Token (aMRT) is proposed as a metric to measure the performance of pronunciation. Formally, 
\begin{equation*}
    \mathbf{aMRT} = \frac{1}{k} \sum_{j=1}^k \sum_{i \in M_j} \delta(z_i, z_i^*)
\end{equation*}
where $M_j \subset [T]$ represents the $j$-th set of indices to be masked, and function $\delta$ is defined as:
\begin{equation*}
    \delta(z, z^*) = 
    \begin{cases}
    0, \quad & z = z^* \\
    1, \quad & z \ne z^*
    \end{cases}
\end{equation*}

A higher aMRT value corresponds to a greater number of mis-recovered tokens and thus a lower quality of pronunciation. To obtain the PCC results between our proposed metrics and ground-truth scores, we adopt the negative values of aMRT as our final metrics.

% The difference between $Z$ and $Z^*$ emerges when mispronunciation components are masked. As such, taking the average contributes to including such cases.

\section{Experiments}

\subsection{Dataset}
We conduct experiments on the dataset speechocean762~\cite{speechocean762}, which is specifically designed for pronunciation assessment. This open-source speech corpus is composed of 5,000 English utterances collected from 250 non-native speakers, half of whom are children. The corpus provides rich label information including phoneme, word, and sentence levels, and includes assessment scores ranging from 0 to 10 annotated by five experts. Our proposed approach is evaluated at the sentence level on the test set, which contains 2,500 utterances. We choose this public dataset for easy reproduction and comparison.

% In this study, we conduct experiments on the dataset speechocean762\cite{speechocean762}, which is specifically designed for pronunciation assessment. Speechocean762 is an open-source speech corpus collected from 250 non-native speakers and half of them are children. This corpus contains 5000 English utterances. We only evaluate our proposed approach on the test set of it, containing 2500 utterances in total. Speechocean762 provides rich label information including phoneme-level, word-level, and sentence-level labels from various aspects. Among them, we only take usage of the total scores of the utterance level ranging from 0-10. Each score is annotated by five experts.  

\subsection{Baseline Models}
We compare our proposed method with regression-based and non-regression-based baselines. The regression-based baselines include GoP~\cite{phonegop2000,gopbi}, DeepFeature\footnote{DeepFuture refers to the methods in~\cite{deeptransfer} using deep features of ASR acoustic model}~\cite{deeptransfer}, and the state-of-the-art GOPT~\cite{gopt}, all of which are supervised with human-annotated pronunciation scores. The non-regression-based baseline, on the other hand, utilizes the average phoneme-level GoP over the entire sentence as the measurement, and is referred to as non-reg GoP. This method does not require score annotations but instead uses a supervised ASR model.

% We compare our proposed method with regression-based and non-regression-based baselines. Regression-based baselines include GoP\cite{phonegop2000}, DeepFeature\footnote{DeepFuture refers to the methods in \cite{deeptransfer} using deep futures of ASR acoustic model}\cite{deeptransfer}, and the state-of-the-art GOPT\cite{gopt}. These regression-based baselines are supervised with human-annotated assessment scores. The non-regression-based baseline here refers to the method using the average phoneme-level GoP over the whole utterance as the measurement, dubbed as non-reg GoP. This baseline does not require score annotations but uses a supervised ASR model.

\subsection{Experimental Setup}

We utilize the HuBERT-Base\footnote{https://github.com/pytorch/fairseq} model and adopt the CNN encoder, Transformer encoder, and k-means clustering in the experiments. HuBERT-Base is pre-trained on the LibriSpeech-960~\cite{librispeech}, and the k-means with 100 clusters is fitted on LibriSpeech train-clean-100 split as per~\cite{softunit} using intermediate representations from HuBERT-Base. The output of the 7th layer of the Transformer Encoder is chosen as the default feature for clustering, as the resulting acoustic units perform well in discrimination tests~\cite{hubert,zeroben,zerospeaker}. We set masking probability $p=20\%$, masking length $l=5$, and repeating times $k=50$ as the default. Each experiment is repeated three times with three different random seeds $\{13, 21, 100\}$, and the mean and standard deviation of the results are reported. Prior to performing the inference steps, all input audios are resampled with 16000 as the sample rate. The non-reg GoP is computed using Kaldi~\cite{kaldi} to obtain the average phoneme-level GoP of the entire sentence. The ASR model utilized in this calculation is Librispeech ASR Chain Model\footnote{https://kaldi-asr.org/models/m13}, as per~\cite{gopt}.

% We utilize the HuBERT-Base model\footnote{https://github.com/pytorch/fairseq} as our backbones for the experiments. HuBERT-Base is pre-trained on the LibriSpeech-960 \cite{librispeech}, and the K-means clustering with 100 clusters inside the model is fitted iteratively. We choose the output of the 7th layer of the transformer encoder as the default feature for clustering. This is because the resulting acoustic units of them have better performance on discrimination tests\cite{hubert,zeroben,zerospeaker}. As for masking, we set masking probability $p=20\%$, masking length $l=5$, and repeating times $k=50$ as default. For each experiment, we run the same experiments three times with three different random seeds and report the mean and standard deviation of the results. Before performing the inference steps, we resample all input audios with 16000 as the sample rate. The PCC scores of given sentence are calculated for evaluation. For the non-reg GoP, we use Kaldi to compute the average phoneme-level GoP of the whole sentence. The ASR model we used is Librispeech ASR Chain Model \footnote{https://kaldi-asr.org/models/m13} following \cite{gopt}.

\subsection{Main Results}
Two comparative studies are conducted to assess the effectiveness of the proposed method. The first study involves PCC performance comparison between our proposed method with regression-based and non-regression-based baselines, while the second study compares the PCC performance of different masking strategies. 

The performances of various regression-based baselines and non-regression-based baselines are presented in Table~\ref{tab:main}. The results indicate that, compared to regression-based baselines, the proposed method lags behind the basic supervised baseline by a small margin of 0.04 PCC, despite a large performance gap of 0.14 PCC with the state-of-the-art supervised baseline. Notably, the proposed method is achieved by leveraging only the acoustic knowledge of HuBERT-Based acquired during pretraining, without the usage of annotated scores. 

Furthermore, in comparison with the non-regression-based baseline, our proposed method shows a performance improvement of 0.03 PCC over the non-reg GoP. It is noteworthy that non-reg GoP requires an ASR model, while our method does not, underscoring the effectiveness of our ASR-free approach.   

% Two comparative studies are conducted to demonstrate the effectiveness: (1) comparison with regression-based and non-regression-based baselines; (2) comparison with different masking strategies 

% Table \ref{tab:main} shows the performance of various regression-based baselines and non-regression-based baselines. Compared with regression-based baselines, despite the large performance gap between our proposed method with SOTA supervised baselines of 0.74 PCC, our proposed method has 0.03 PCC lagging behind the basic supervised baseline of 0.64 PCC. This finding is promising since we only leverage the acoustic knowledge of HuBERT-Based learned during pretraining, without the usage of annotated scores. Furthermore, compared with the non-regression-based baseline, our proposed method outperforms the non-reg GoP by 0.04 PCC. Notice that non-reg GoP needs an ASR model while ours does not, which demonstrates the effectiveness of our ASR-free and training-free method. 

\begin{table}[htbp]
  \caption{Comparison between our method with regression-based and non-regression-based baselines on speechocean762}
  \label{tab:main}
  \centering
  \begin{tabular}{ cccccccccccc }
    \toprule[1pt]
    \multicolumn{6}{c}{\textbf{Model}} & \multicolumn{6}{c}{\textbf{PCC}} \\
    \hline
    \multicolumn{6}{c}{\textbf{Regression based}} & \multicolumn{6}{c}{}  \\
    \hline
    \multicolumn{6}{c}{GoP~\cite{phonegop2000}}             &      \multicolumn{6}{c}{0.64}             \\
    \multicolumn{6}{c}{GoP(2BLSTM+MLP)~\cite{gopbi}}      &      \multicolumn{6}{c}{0.67}                 \\
    \multicolumn{6}{c}{DeepFeature~\cite{gopfeat}}          &      \multicolumn{6}{c}{0.72}                \\
    \multicolumn{6}{c}{GOPT~\cite{gopt}}                 &      \multicolumn{6}{c}{0.74}                \\

    \hline
    \multicolumn{6}{c}{\textbf{Non-regression based}} & \multicolumn{6}{c}{}  \\
    \hline
    \multicolumn{6}{c}{non-reg GoP}              &       \multicolumn{6}{c}{0.57}     \\
    \multicolumn{6}{c}{Ours}                &       \multicolumn{6}{c}{0.60}     \\
    \bottomrule[1pt]
  \end{tabular}
\end{table}

Table~\ref{tab:mask_strategy} presents the performance comparison of two masking strategies employed in this study. The results show that random masking achieves superior performance with an improvement of 0.014 PCC over regular masking. We conjecture that this may be due to the fact that the input distribution with random masking is closer to the input distribution during pretraining, leading to enhanced performance. In addition, the experimental results reveal that random masking exhibits a low variance, indicating the stability of the method.

% Table \ref{tab:mask_strategy} displays the performances of two masking strategies we employed. It is observed that random masking outperforms regular masking by 0.03 PCC. We conjecture that the reason comes from the masking way during pretraining. The input with random masking is closer to the input distribution of pretraining. Besides, we discover that random masking has a low variance of
% 0.002, indicating that random masking is stable as well. 

\begin{table}[htbp]
  \caption{Comparison of two masking strategies. The standard derivation of Random Masking is reported. }
  \label{tab:mask_strategy}
  \centering
  \begin{tabular}{ cccccccccccc }
    \toprule[1pt]
    \multicolumn{6}{c}{\textbf{Masking Strategy}} & \multicolumn{6}{c}{\textbf{PCC}} \\
    \hline
    \multicolumn{6}{c}{Random Masking}       &      \multicolumn{6}{c}{$0.595 \pm 0.002$}         \\
    \multicolumn{6}{c}{Regular Masking}      &      \multicolumn{6}{c}{$0.581 $}          \\

    \bottomrule[1pt]
  \end{tabular}
\end{table}

\subsection{Impact of masking hyperparameters}
% analysis on the impact of masking probability, masking length, and feature layers that are used for clustering

\subsubsection{Random Masking}
\begin{figure*}[htbp]
  \centering
  \subfigure[]{
      \label{fig:a}
      \includegraphics[scale=0.22]{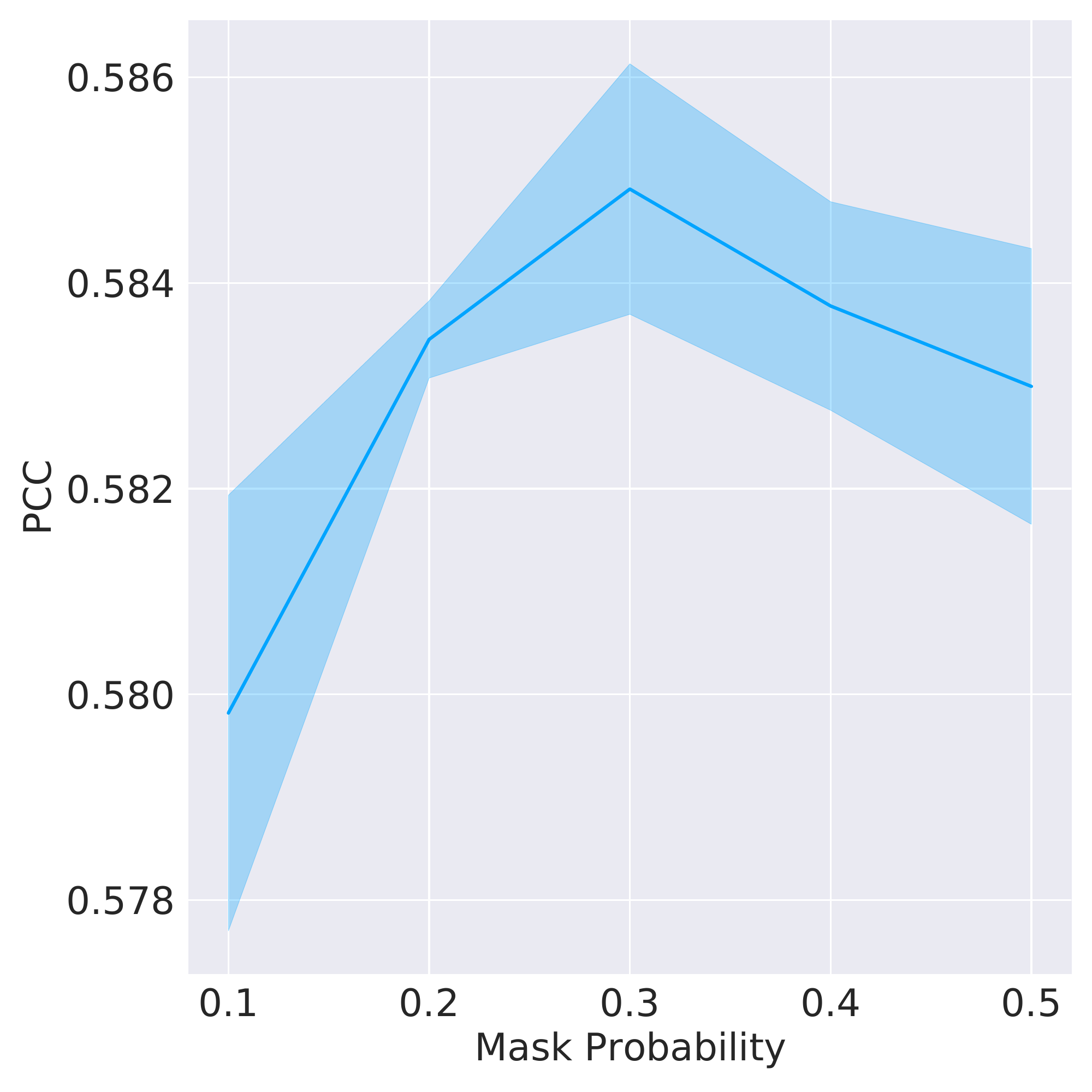}
  }\hspace{-3mm}
  \subfigure[]{
      \label{fig:b}
      \includegraphics[scale=0.22]{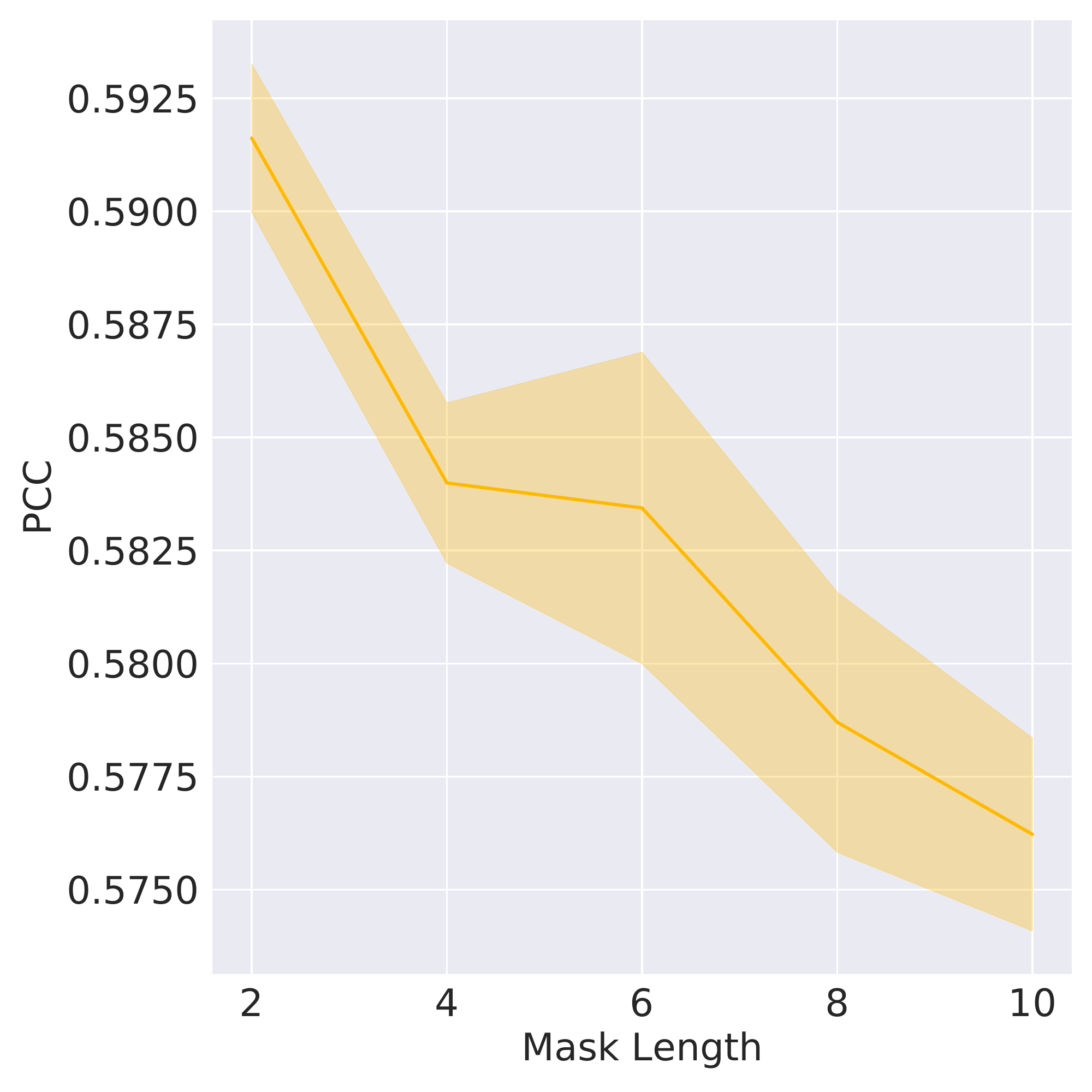}
  }\hspace{-3mm}
  \subfigure[]{
      \label{fig:c}
      \includegraphics[scale=0.22]{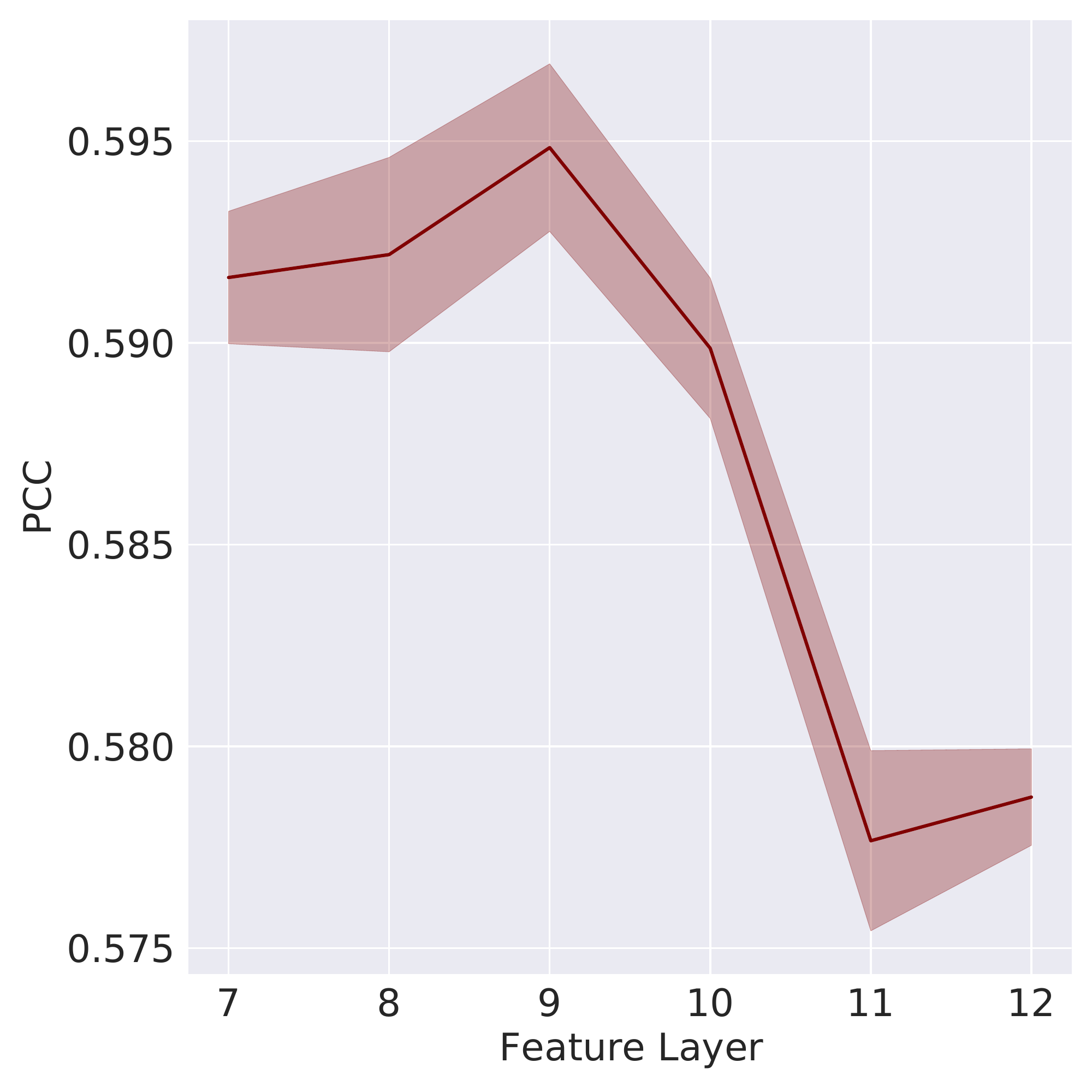}
  }
  \vspace{-10pt}
  \caption{Impact of (a) mask probability, (b) mask length, and (c) feature layer on PCC results}
  \label{ablation}
  \vspace{-10pt}
\end{figure*}

In order to further examine the impact of various hyperparameters of random masking, including masking probability, masking length, and feature layers used for clustering on the final results, three additional experiments are carried out. The results are presented in Figure~\ref{ablation}. 

The initial subfigure~\ref{fig:a} illustrates the impact of mask probability on PCC results, with mask probability ranging from 0.1 to 0.5 with an interval of 0.1. The mask length is set to 5, and the feature layer is set to 7. The results indicate that the mask probability of 0.3 yields the best performance, while both higher and lower mask probabilities produce inferior outcomes. This observation may be attributed to the fact that the high mask probability may discard essential information that is required for reconstruction, whereas the low mask probability may decrease the possibility of masking mispronunciation parts.

Subfigure~\ref{fig:b} showcases how the length of each masked span affects the PCC results. The mask length ranges from 2 to 10 with an interval of 2, while the mask probability is set to 0.2, and the feature layer is set to 7. The curve of this figure suggests a linear decrease in performance as the length increases. This phenomenon may stem from the pre-trained HuBERT-Base's inadequate ability to recover a long masked span given the context. 

Apart from the aforementioned factors, this study also investigates the degree to which the features used for clustering can contribute to pronunciation assessment. Therefore, the features from various layers of the Transformer Encoder ranging from 7 to 12 are examined. The outcomes presented in subfigure~\ref{fig:c} reveal that using features from the 9th layer results in the best PCC performance. Generally, features from the 7th to 10th layer are considered useful for pronunciation assessment, whereas deeper features lead to poorer performance.

\subsubsection{Regular Masking}
\begin{figure}[htbp]
  \centering
  \includegraphics[scale=0.25]{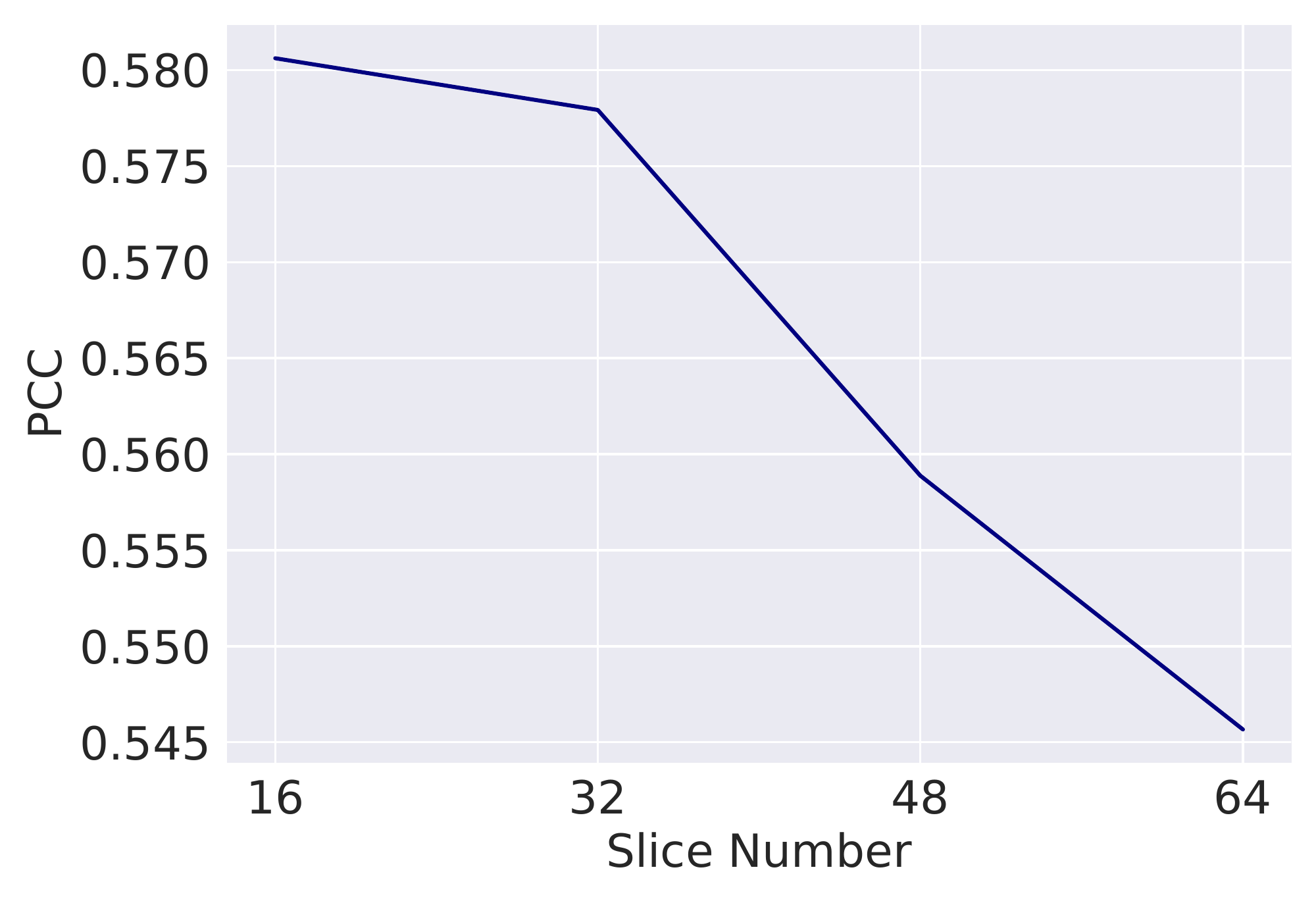}
  \vspace{-10pt}
  \caption{Impact of slice number on PCC results}
  \label{ablation_reg}
  \vspace{-10pt}
\end{figure}

For regular masking, we mainly investigate the impact of the slice number on the PCC results, namely how the mask granularity affects the PCC results. The results are presented in Figure~\ref{ablation_reg}. Our finding suggests that the more refined granularity of a single mask span does not necessarily lead to improved performance. One potential explanation for this outcome is that the use of a single mask span causes a shift from the input distribution, leading to poor performance. In addition, shorter masked spans may fail to cover entire words or even phonemes, which can have an adverse impact on the results. 

\section{Discussion}
While our zero-shot method achieves results comparable to supervised methods, it is essential to acknowledge that our method differs from the canonical-text-based pronunciation assessment. Our method draws on the acoustic knowledge obtained during pre-training, and thus, even if the transcription is different from the canonical text, a speech that is accurately pronounced may still receive a high score. Moreover, our method is limited to sentence-level assessment, and the exploration of unsupervised pronunciation assessment at the phoneme and word levels will be left as future work. The objective of this study is to establish a baseline and provide a pilot study of unsupervised pronunciation assessment.

% Even though our proposed zero-shot method achieves performance close to supervised counterparts, we need to point out that our method 
% does not apply to cases when grounded texts are provided for assessing pronunciation. This is a key difference from other work. We leverage the acoustic knowledge of SSL pre-trained acoustic models for the assessment. Therefore, a well-pronounced speech with totally different semantics from the ground text may be also scored high. Besides, our method allows utterance-level assessment. We leave the exploration of unsupervised phoneme-level and word-level pronunciation assessment as future work. For this work, we hope to establish the unsupervised baselines of pronunciation and provide a pilot study in the pronunciation assessment community.

\section{Conclusion}

In this paper, we present a zero-shot automatic pronunciation assessment approach. Instead of training regression models or using ASR models to compute GoP, we directly utilize a SSL pre-trained acoustic model and use the acoustic knowledge acquired from pre-training. To perform the ASR-free pronunciation assessment, we design two masking strategies and a novel evaluation metric to score the pronunciation of given speeches at the sentence level. Experimental results on speechocean762 achieve comparable performance to the supervised regression-based baseline and outperform the non-regression-based baseline. In the future, we hope to extend this research line of unsupervised pronunciation assessment to phoneme and word levels.

% In this work, we propose a zero-shot pronunciation assessment approach that requires no labeled data. This is achieved by leveraging the SSL pre-trained acoustic model, HuBERT \cite{hubert}, for conducting the masked token prediction task that is performed during pre-training. We convert input speech frames into token sequences using k-means clustering and mask a portion of the tokens. Then, we feed the masked sequences into the frozen HuBERT and utilize HuBERT to recover masked spans given the contexts. Finally, a scoring module, evaluating the gap between the output of HuBERT and the input sequences, is designed to measure the pronunciation of a given speech. Our proposed method is unsupervised and training-free. We conduct experiments on the speechocean762 dataset\cite{speechocean762}. The experimental results demonstrate the effectiveness of our proposed method in terms of the Pearson Correlation Coefficient.

\section{Acknowledgements}
The authors would like to thank anonymous reviewers for their valuable
suggestions. This project is funded in part by a research grant MOESOL-2021-0017 from the Ministry of Education in Singapore.

% As a final reminder, the 5th page is reserved exclusively for references. No other content must appear on the 5th page. Appendices, if any, must be within the first 4 pages. The references may start on an earlier page, if there is space.

\newpage

\bibliographystyle{IEEEtran}
\bibliography{mybib}

\end{document}